# *Collaborative Trust*: A Novel Paradigm of Trusted Mobile Computing

Tatini Mal-Sarkar[*]  
Hathaway Brown High School  
Cleveland, OH  
tmal-sarkar13@hb.edu

Swarup Bhunia  
Case Western Reserve University  
Cleveland, OH  
swarup.bhunia@case.edu

With increasing complexity of modern-day mobile devices, security of these devices in presence of myriad attacks by an intelligent adversary is becoming a major issue. The vast majority of cell phones still remain unsecured from many existing and emerging security threats [1]. There are two major modes of threats for mobile wireless devices - hardware and software. In the hardware field, many threats exist. Such threats include vulnerability to system malfunction, weakness of authentication procedures, and increasing threat of hardware Trojans [2-10]. In software, these threats include electronic eavesdropping, location tracking software, and malware [1] (software Trojan horses and viruses). In general, the security attacks in mobile devices – through either hardware or software or combination of both – primarily aim at either leaking personal information or causing system malfunction.

*To address the security threats in mobile devices we are exploring a technology, which we refer as "Collaborative Trust". It is a technology that uses a system of devices cooperating with each other (working in a fixed or ad-hoc network) to achieve the individual security of each device. The idea is that each device is insecure by itself, since in many cases it is incapable of checking its safety by itself (e.g. when it is compromised it may lose its ability to monitor its own trustworthiness), but together, they can ensure each other's security in a collaborative manner.* Let us consider a group of five mobile wireless devices, as shown in the adjacent figure. Each device can perform a specific function and check the results of this function computed by four other devices. If output of a function generated by the $i^{th}$ device is not agreed by majority of the remaining devices then we can infer that the $i^{th}$ device

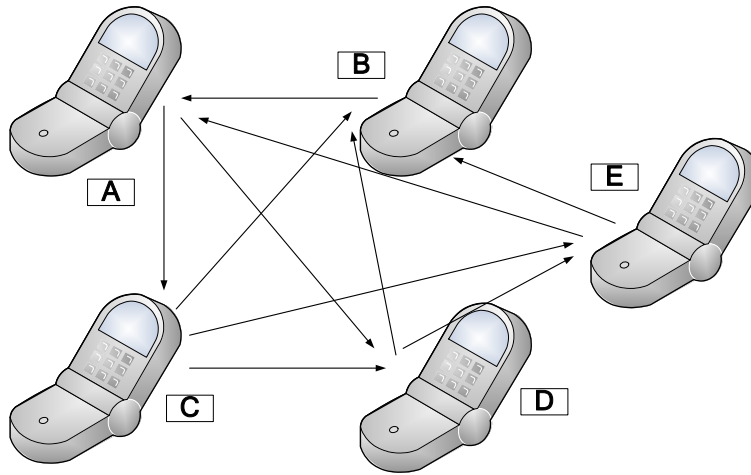

A collaborative network of devices for checking security of individual devices in cooperative manner.

is malfunctioning. The hypothesis holds when an attack is manifested in the output of the specific function and when all the devices in the collaborative network are not affected at the same time.

The technology of collaborative trust can be implemented as an "application" onto smart phones, such as the iPhone. It would be able to check for different security threats, e.g. threats of hardware Trojans [2-10]. An attacker can mount a hard-to-detect hardware Trojan attack, which can evade manufacturing test can validation [5-6]. With increasingly complex hardware design in case of mobile devices, these devices are becoming increasingly vulnerable to hard-to-detect malicious hardware modifications. Runtime monitoring of device functionalities using the proposed collaborative approach can be an effective solution to ensure trusted field operation in presence of hardware Trojan threats. The "app" would essentially run a set of test routines, for example add, multiply, comparison, or more complex routines which combine number of functions, and compare the function outputs from one device with those of number of other devices. The scheme can be fully implemented as software on top of the operating system (OS) kernel (which is assumed to be trusted) and hence, would not require any additional hardware insertion or modification of existing hardware. The figure above shows a flow chart to illustrate part of a possible sequence of activities that each device would be required to do to realize the proposed collaborative trust scheme.

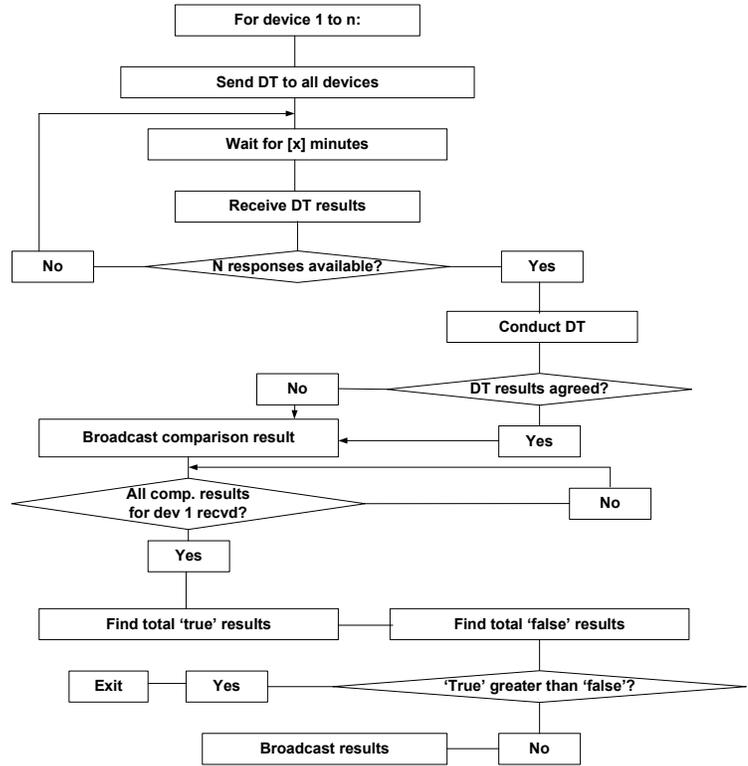

Flowchart showing part of a possible sequence of activities for each device in a collaborative trust network.

Though the proposed technology can potentially address many of today's security issues, there are several challenges to address. 1) It is important that the attacker cannot affect all the devices in a collaborative network simultaneously. If the five-member group described above is formed in an ad-hoc manner, an attacker would have no idea which devices belong to which system. Thus, to affect the system in a way that bypasses the collaborative trust scheme would be extremely difficult unless the adversary corrupts more than half the number of devices at a time. 2) The test routines need to be designed such that they can manifest the effect of a security attack e.g. would be able to trigger a hardware Trojan. It should comprise of a set of small routines which can try to comprehensively cover different attack modes and incur minimum performance and energy overhead. Minimizing the impact on battery life, while providing high assurance, would be of high importance. 3) Developing a communication protocol that allows a collection of device to autonomously check the security of individual devices and take appropriate action in case a threat is detected in one or more device will be an important challenge. The devices need to be capable of initiating the task of checking other devices; receiving the results of test routines back; comparing and sending the result of comparison to neighboring devices; and accepting the check request from other devices in the collaborative network.